\documentclass[11pt]{article}

\usepackage{graphicx}
\usepackage{epstopdf}
\usepackage{amsmath}
\usepackage{amssymb}
\usepackage{amsfonts}
\usepackage{amsthm}
\usepackage[usenames]{color}
\usepackage{array}
\usepackage{physics}

\usepackage[
colorlinks=true,
linkcolor=blue,
urlcolor=blue,
filecolor=blue,
citecolor=red,
pdfstartview=FitV,
pdftitle={},
pdfauthor={},
pdfsubject={},
pdfkeywords={},
pdfpagemode=None,
bookmarksopen=true
]{hyperref}

\usepackage{epsfig}

\usepackage{hyperref}

\newcommand{\bea}{\begin{eqnarray}}
\newcommand{\eea}{\end{eqnarray}}
\newcommand{\ba}{\begin{array}}
	\newcommand{\ea}{\end{array}}
\newcommand{\ee}{\end{equation}}

\numberwithin{equation}{section}

\begin{document}

\begin{flushright}
	\texttt{\today}
\end{flushright}

\begin{centering}
	
	\vspace{2cm}
	
	\textbf{\Large{
			  Holographic CCFT Pseudo-Entropy }}
	
	\vspace{0.8cm}
	
	{\large   Reza Fareghbal, Abolfazl Hassani Majoulan }
	
	\vspace{0.5cm}
	
	\begin{minipage}{.9\textwidth}\small
		\begin{center}
			
			{\it  Department of Physics,
				Shahid Beheshti University, 1983969411,
				 Tehran , Iran \\}
				  			
			\vspace{0.5cm}
			{\tt   r$\_$fareghbal@sbu.ac.ir, reza.fareghbal@gmail.com, abolfazl.hassani2016@gmail.com}
			\\
			
					\end{center}
	\end{minipage}


	\begin{abstract}
	According to the flat/CCFT  correspondence, Carrollian conformal field theories (CCFT) in d dimensions are dual to asymptotically flat spacetimes in d+1 dimensions. In this paper, starting from the holographic interpretation of pseudo-entropy in the (A)dS$_3$/CFT$_2$, we show that both extremal spacelike and timelike curves possess a well-defined flat-space limit. The length of these curves can be regarded as the real and imaginary parts of the pseudo-entropy for the underlying field theory, where only the real part has been considered thus far. Our calculations can confirm that the entanglement entropy in the CCFTs is fundamentally pseudo-entropy.

	\end{abstract}

\end{centering}

\newpage



\section{Introduction}
The study of holography for asymptotically flat spacetimes is one of the highly active branches in high-energy physics. One of the candidates for the holographic dual of asymptotically flat spacetimes is provided within the framework of the flat/CCFT correspondence \cite{Bagchi:2010zz, Bagchi:2012cy}. According to this correspondence, Carrollian conformal field theories (CCFT) in $d$ dimensions are dual to asymptotically flat spacetimes in $d+1$ dimensions \cite{Bagchi:2012cy}. CCFTs are essentially ultra-relativistic field theories that are derived from conformal field theories in the zero limit of light speed \cite{leblond}-\cite{Duval:2014uva}. These theories in two and three dimensions possess infinite-dimensional symmetries, whose symmetry group is identical to the asymptotic symmetries at the null infinity of asymptotically flat spacetimes in three and four dimensions. These asymptotic symmetries were introduced by Bondi, van der Burg, Metzner, Sachs (BMS) in the late 1960s \cite{Bondi:1962px, Sachs:1962zza} and have been studied more precisely later in  \cite{Barnich:2006av, Barnich:2009se} . Therefore, we can say that CCFTs are field theories that possess BMS symmetries, and for this reason, they have also been named BMSFTs \cite{Jiang:2017ecm}.

The infinite-dimensional symmetries of CFTs in two and three dimensions yields  universal properties for them, such as the form of n-point functions or the entanglement entropy of subsystems. In this regard, extensive work has been done by various groups, a list of which can be found in the references of the article \cite{Bagchi:2025vri}. According to the flat/CCFT correspondence , all calculations in field theory and all universal properties of CFTs must have a holographic interpretation in the framework of calculations related to asymptotically flat space-times. In fact, the dictionary of flat/CCFT correspondence can be completed through two paths. In the first path, one can directly seek gravitational calculations in asymptotically flat space-times that correspond to the properties of CCFTs. The second path involves starting from AdS/CFT and taking the limit of the calculations. In this method, the flat space limit, which relates the metrics of asymptotically AdS spacetimes to asymptotically flat metrics (the zero limit of the cosmological constant or the infinite limit of the AdS radius), is assumed to be equivalent, on the field theory side, to the ultra-relativistic  limit of the conformal field theory \cite{Bagchi:2012cy}. Using the second method, for any valid calculation within the framework of the AdS/CFT correspondence, an equivalent can be found in the flat/CCFT correspondence, thereby completing the related dictionary.

 One of the interesting calculations in AdS/CFT is the holographic expression for the entanglement entropy of subsystems in CFT, according to which this entropy in the field theory is equivalent to computing the area of extremal codimension-2 spacelike surfaces in asymptotically AdS spacetimes \cite{Ryu:2006bv}. This correspondence has attracted much attention  in recent years, and numerous works on this topic can be found in the literature.
 
 Entanglement entropy can be defined for subsystems in CCFTs, expected to have universal formulas independent of the  details of the theory due to the infinite-dimensional symmetries of CCFT$_2$ and CCFT$_3$. This problem was first studied in \cite{Bagchi:2014iea}  for CCFT$_2$, where entanglement entropy for spacelike intervals in these theories was introduced. The related gravitational calculations, involving the identification of extremal surfaces in asymptotically flat spacetimes, were also conducted in \cite{Jiang:2017ecm}. These extremal surfaces have a fundamental difference from those introduced in AdS/CFT: unlike extremal surfaces in asymptotically AdS spacetimes, they do not connect to the two endpoints of subsystems at the null infinity. Thus, the question of how extremal surfaces in AdS relate to those in flat spacetimes could be an intriguing one.
 
 It is not difficult to see that in suitable coordinates where  flat-space limit is well-defined (BMS coordinate \cite{Barnich:2012aw}), the spacelike extremal surfaces in AdS do not have a well-defined flat-space limit, and thus, this method can not be used to obtain spacelike extremal surfaces in asymptotically  flat spacetimes. Therefore, the answer to our question is not trivial and requires more detailed investigation. We began studying  this problem  in  paper \cite{Fareghbal:2020dtq}, in which we introduced  new spacelike extremal curves in AdS$_3$ whose flat-space limit leads to the curves in \cite{Jiang:2017ecm}: If in BMS coordinates (where  the flat space limit of asymptotically AdS space-times  are well-defined), we denote the boundary of global AdS with $u$ and $\phi$ , where $u$ is retarded time and $\phi$ is a periodic coordinate, then one can represent an arbitrary interval on the boundary as $-{\ell_u\over 2}<u<{\ell_u\over 2},-{\ell_\phi\over2}<\phi<{\ell_\phi\over2}$, where $\ell_u$ and $\ell_\phi$ are constants. By applying an appropriate condition on these constants, the interval can be defined as spacelike. The curves obtained using the RT proposal are connected to both ends of this interval, and their equations do not have a well-defined flat space limit.Our observation in \cite{Fareghbal:2020dtq} was that there exist new curves obtained from RT curves through the transformation $\ell_u\leftrightarrow \ell\ell_\phi$ (where $\ell$ is the AdS radius), and their flat-space limit is well-defined, leading to  curves of \cite{Jiang:2017ecm}. These new curves connect to the AdS boundary at new points $\left(u=\pm{\ell\ell_\phi\over 2},\phi=\pm{\ell_u\over 2 \ell}\right)$ that are the two ends of a timelike interval.

Recently, in \cite{Doi:2022iyj, Doi:2023zaf}, entanglement entropy for timelike intervals or pseudo-entropy was introduced. Using the definitions of \cite{Doi:2022iyj}, we realized in \cite{Fareghbal:2024lqa} that the new curves that have a well-defined flat-space limit are actually the spacelike portions of the curves obtained in \cite{Doi:2022iyj} for the holographic dual of pseudo-entropy. The question left unanswered in \cite{Fareghbal:2024lqa} was what happens to timelike curves, whose lengths are proportional to the imaginary part of pseudo-entropy, after taking the flat-space limit.

In this paper, we demonstrate that not only  the space-like curves related to pseudo-entropy have a well-defined flat-space limit, but their timelike component also has a well-defined limit, leading to timelike curves in three-dimensional flat spacetime. Lengths of these new timelike curves can contribute an imaginary term to the entanglement entropy of dual CCFTs. Therefore, our claim in this article is that the entanglement entropy of CCFTs should fundamentally be pseudo-entropy, the imaginary part of which we introduce in this paper using flat/CCFT holography. 

The dictionary of flat/CCFT should be obtained not only by taking the flat-space limit from AdS/CFT but also by using holography of asymptotically dS spacetimes.   We show that the holographic dual for the pseudo-entropies, introduced in  \cite{Doi:2022iyj} (see also \cite{Narayan:2022afv, Narayan:2023zen}) and possessing both timelike and spacelike curves, has a well-defined flat-space limit. These curves, after taking the flat-space limit, precisely correspond to the curves obtained from the limit of timelike-entropy curves in AdS spacetime. 

\section{Preliminaries}
\subsection{Entanglement entropy and pseudo-entropy in quantum field theory}\label{pseudo t}

The concept of entanglement entropy arises from a fundamental observation in quantum mechanics. For a pure state, if the total Hilbert space ${\mathcal{H}}$ is decomposed into two subsystems ${\mathcal{H}}={\mathcal{H}}_A\otimes {\mathcal{H}}_B$, the overall state of the system is not necessarily expressible as a tensor product of the states of these subsystems. This non-factorizability gives rise to quantum entanglement, for which the von Neumann entropy can be used as a quantitative measure.

Let $\ket{\psi}$ be a pure state and $\rho=\ket{\psi}\bra{\psi}$ denote the total density matrix of the system. The information accessible to an observer confined to the region $A$, which is a spacelike region, is described by the reduced density matrix
\begin{equation}\label{rhoa}
\rho_A=\text{Tr}_B \rho.
\end{equation}
The entanglement entropy for subsystem $A$ is defined as the  von Neumann entropy of $\rho_A$,
\begin{equation}\label{SA}
S_A=-\text{Tr}\left[\rho_A\log \rho_A\right].
\end{equation}
Physically, this quantity measures the amount of information lost by observer $A$ due to the separation from subsystem $B$, or equivalently, it quantifies the degree of entanglement between parts $A$ and $B$.

A direct computation of this quantity is highly nontrivial due to the presence of $\log \rho_A$. However, powerful techniques such as the replica method  make such calculations feasible:
\begin{align}\label{reni}
\nonumber & S_A^{(n)} = \frac{1}{1 - n} \ln \mathrm{Tr}_A \left( \rho_A^n \right),\\
&S_A = \lim_{n \to 1} S_A^{(n)} = - \mathrm{Tr}_A \left( \rho_A \ln \rho_A \right).    
\end{align}
In fact, instead of computing expression~\eqref{SA} directly, one evaluates expression~\eqref{reni} in the $n \to 1$ limit of the Rényi entropy.  For CFT$_2$ with infinite-dimensional symmetry, applying this method leads to universal formulas for the entanglement entropy \cite{Holzhey:1994we}-\cite{Calabrese:2009qy}.

The concept of pseudo-entropy involves generalizing the density matrix to a more abstract mathematical object. To this end, let us consider two distinct pure states, $\ket{\psi}$ and $\ket{\varphi}$. The transition matrix for the entire system is defined as \cite{Nakata:2020luh}
\begin{equation}
\tau=\frac{\ket{\psi}\!\bra{\varphi}}{\bra{\varphi}\ket{\psi}}.
\end{equation}
The reduced transition matrix for subsystem $A$ is obtained by tracing over the degrees of freedom associated with subsystem $B$:
\begin{equation}\label{tranmatrix}
\tau_A=\text{Tr}_B\left[\frac{\ket{\psi}\!\bra{\varphi}}{\bra{\varphi}\ket{\psi}}\right].
\end{equation}
The pseudo-entropy is then defined as the von Neumann entropy of this operator:
\begin{equation}\label{Sp}
S_A^{(p)}=-\text{Tr}\left[\tau_A\log \tau_A\right].
\end{equation}
Due to the non-Hermitian nature of the transition matrix, the pseudo-entropy is generally a complex quantity, and its physical interpretation is not as straightforward as that of the standard entropy. Nevertheless, it plays several key roles in various physical contexts.

\begin{itemize}
    \item \textbf{Time-like intervals:} When subsystem $A$ corresponds to a timelike region, the standard definition of the density matrix leads to a non-Hermitian operator. In this case, the entropy computed from expression~\eqref{SA} acquires a complex value. This timelike entanglement entropy can therefore be interpreted precisely as a form of pseudo-entropy.

    \item \textbf{Non-unitary theories:} In a non-unitary theory, the standard Hermitian structure (i.e., conjugation) is modified. Consequently, the operator $\rho_A$ is no longer Hermitian and instead behaves similarly to a transition matrix. Therefore, what appears to be a standard entanglement entropy is, in fact, a pseudo-entropy. 
\end{itemize}

\subsection{The Flat/CCFT correspondence}
The asymptotic symmetries at null infinity in asymptotically flat spacetimes in three and four dimensions are infinite-dimensional and are called BMS symmetries  \cite{Barnich:2006av, Barnich:2009se}. In three dimensions, these symmetries are given by the following algebra \cite{Barnich:2006av}:
\begin{align}\label{algebra}
\nonumber & [L_n,L_m]=(n-m)L_{n+m}+\frac{c_L}{12}(n^3-n)\delta_{n+m,0},\\
\nonumber & [L_n,M_m]=(n-m)M_{n+m}+\frac{c_M}{12}(n^3-n)\delta_{n+m,0},\\
& [M_n,M_m]=0,
\end{align}
where $c_L$ and $c_M$ are central charges.

If we want to propose a duality analogous to AdS/CFT for asymptotically flat spacetimes, one suggestion could be to look for a field theory in one dimension lower than the asymptotically flat space, which possesses symmetry \eqref{algebra} \cite{Bagchi:2010zz}. In fact, reference , \cite{Bagchi:2012cy} observed that these symmetries are obtained from the ultra-relativistic limit (c $\to$ 0) of a conformal field theory of the same dimension. Therefore, these theories, which are dual to asymptotically flat spacetimes in one higher dimension, are called CCFT, and this duality is also named flat/CCFT.

 As implied by \eqref{algebra}, a two-dimensional CCFT, analogous to a two-dimensional CFT, exhibits infinite-dimensional symmetries. The presence of these symmetries allows for the derivation of a universal formula for entanglement entropy in such field theories. This derivation  were carried out in \cite{Bagchi:2014iea} , where it was shown that for a two-dimensional CCFT defined on a cylinder with coordinates $u$ and $\phi$, with $\phi$ a periodic spatial coordinate and $u$ the retarded time coordinate and  $A$ is a region on the cylinder, defined by 
$\left(u=-\frac{l_u}{2}, \phi=-\frac{l_\phi}{2}\right)$ and 
$\left(u=\frac{l_u}{2}, \phi=\frac{l_\phi}{2}\right)$, 
where $l_u$ and $l_\phi$ are constants, the entanglement entropy is given by
\begin{equation}\label{ee cylinder}
S_{EE}=\frac{c_L}{6}\log\left(\frac{2}{\epsilon}\sin\frac{l_\phi}{2}\right)+\frac{c_M}{12}l_u\cot\frac{l_\phi}{2}.
\end{equation}
The holographic interpretation of \eqref{ee cylinder} was proposed in \cite{Jiang:2017ecm} as the length of some extremal spacelike curves within the asymptotically flat spacetimes.

\section{Flat-space limit of holographic pseudo-entropy in AdS spacetime}
\subsection{Extremal curves and their flat-space limit}
In this section, using the  proposal of \cite{Doi:2022iyj, Doi:2023zaf}, we aim to find the holographic dual in three-dimensional AdS spacetime for the entanglement entropy corresponding to a timelike interval in the dual CFT, and then take the flat-space limit of it. Taking the flat-space limit is not possible in every coordinate system; typically, it is necessary to perform the calculations in BMS coordinates \cite{Barnich:2012aw}. On the other hand, the calculation related to the holographic dual of entanglement entropy is more easily carried out in Poincaré coordinates, and the results of \cite{Doi:2022iyj, Doi:2023zaf} can be utilized. Therefore, we will also start with AdS$_3$ in Poincaré coordinates, write down the geodesics corresponding to the holographic dual of the entanglement entropy in these coordinates, then transform them to BMS coordinates, and finally take the flat-space limit.

\begin{figure}[h]
\centering\includegraphics[scale=1,width=70mm]{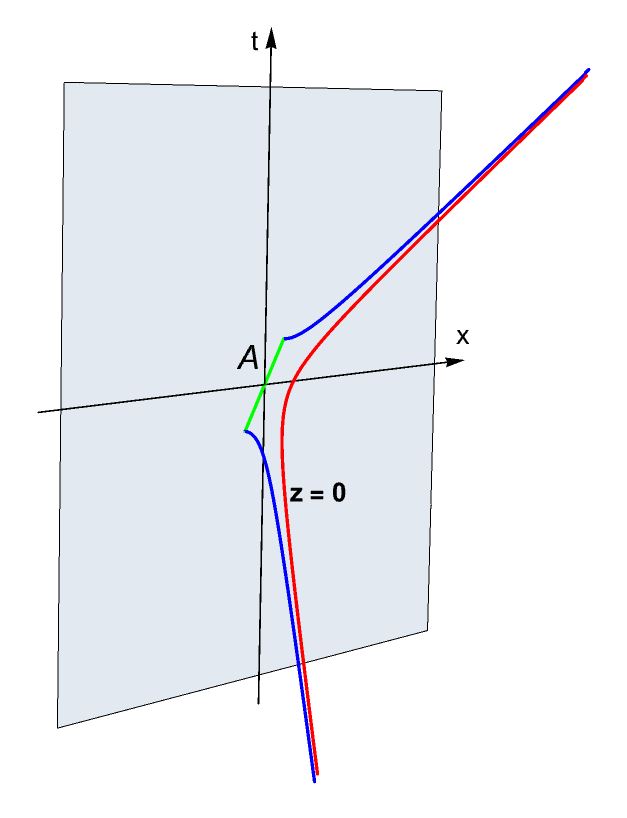}
\caption{ Spacelike (blue) and timelike (red) extremal curves in Poincaré coordinates which are holographic dual of timelike entanglement entropy of timelike interval $A$ (green line) in the boundary CFT.}\label{poincare fig}
\end{figure}

The AdS$_3$ metric in Poincaré coordinates is given by 
\begin{equation}\label{Poincare}
ds^2=\dfrac{\ell^2}{z^2}\left(-dt^2+dx^2+dz^2\right),
\end{equation}
 where $\ell$ is the AdS radius, $z$ is the radial coordinate, and the boundary is located at $z=0$. The boundary coordinates are denoted by $t$ and $x$. We consider a timelike interval with the following specifications on the boundary of the above metric:
\begin{equation}\label{interval timelike}
-\dfrac{l_x}{2}<t<\dfrac{l_x}{2},\qquad -\dfrac{l_t}{2}<x<\dfrac{l_t}{2}.
\end{equation}
$l_x$ and $l_t$ are two constants, which, assuming $l_x > l_t$, will constitute a timelike interval. This interval describes a subsystem in the dual $\text{CFT}_2$ to the $\text{AdS}_3$ spacetime. The entanglement entropy for this timelike subsystem is inherently a complex quantity, and is therefore called the pseudo-entropy. According to the proposal of \cite{Doi:2022iyj, Doi:2023zaf}, the real and imaginary parts of the pseudo-entropy are proportional to the lengths of the spacelike and timelike geodesics, respectively, in the $\text{AdS}_3$ spacetime. The spacelike geodesic (or curve) connects to the two endpoints of the interval $\left(u=\pm\frac{\ell l_\phi}{2},\phi=\pm\frac{l_u}{2\ell}\right)$ on the boundary. This curve is actually composed of two branches, with each branch connecting to one endpoint of the interval. The timelike curves, whose lengths are proportional to the imaginary part of the pseudo-entropy, connect to the other ends of the branches of the spacelike curve, as shown in figure \ref{poincare fig}. For the given interval \eqref{interval timelike}, the spacelike curve is given by
\begin{align}\label{S-L in poincare}
\nonumber x^2=\dfrac{l_t^2\left(4z^2+l_x^2-l_t^2\right)}{4\left(l_x^2-l_t^2\right)},\\
t^2=\dfrac{l_x^2\left(4z^2+l_x^2-l_t^2\right)}{4\left(l_x^2-l_t^2\right)},
\end{align}
 and the timelike curve is given by the following equations:
 \begin{align}\label{T-L in poincare}
\nonumber x^2=\dfrac{l_t^2\left(4z^2-l_x^2+l_t^2\right)}{4\left(l_x^2-l_t^2\right)},\\
t^2=\dfrac{l_x^2\left(4z^2-l_x^2+l_t^2\right)}{4\left(l_x^2-l_t^2\right)}.
\end{align}

From \eqref{S-L in poincare} and \eqref{T-L in poincare}, it is clear that the timelike curve is stretched between $\dfrac{\sqrt{l_x^2-l_t^2}}{2}<z<\infty$, while the spacelike curve is stretched across the entire $\text{AdS}_3$ space in Poincaré coordinates\footnote{With the  transformation $l_x\leftrightarrow l_t$, the spacelike curve \eqref{S-L in poincare} transforms into a spacelike one  whose length is proportional to the entanglement entropy associated with the spacelike interval $-\dfrac{l_x}{2}<x<\dfrac{l_x}{2}, -\dfrac{l_t}{2}<t<\dfrac{l_t}{2}$. In contrast to the spacelike curve \eqref{S-L in poincare}, this spacelike curve is single-branched and connects the two endpoints of the interval.}. (figure \ref{poincare fig})

It is evident from  metric \eqref{Poincare} and   equations \eqref{S-L in poincare} and \eqref{T-L in poincare} that the flat space limit or the limit of $\ell \to \infty$ is not well-defined in the Poincare coordinates. Therefore, we define another  coordinate, known as BMS coordinates, as follows:
\begin{align}\label{coordinate change}
\nonumber z&=\dfrac{2\ell^2}{\Delta},\\
\nonumber t&=\dfrac{2\ell\left(r\sin{\dfrac{u}{\ell}}-\ell\cos{}\dfrac{u}{\ell}\right)}{\Delta},\\
\nonumber x&=\dfrac{2\ell r \sin{\phi}}{\Delta},\\
 \Delta&=\ell\sin{\dfrac{u}{\ell}}+r\cos{\dfrac{u}{\ell}}+r\cos{\phi}.
 \end{align}
In the new coordinates, $u$ is the retarded time and $\phi$ is periodic with a period of $2\pi$. The boundary is also located at $r=\infty$. In these coordinates, the AdS$_3$  metric is written as 
\begin{equation}\label{BMS-global}
  ds^2=-\left(1+\dfrac{r^2}{\ell^2}\right)du^2-2dudr+r^2d\phi^2.
  \end{equation} 
This metric is known as AdS$_3$ metric in global BMS coordinate. If we relate the constants $l_u$ and $l_\phi$ to the constants $l_x$ and $l_t$ as 
\begin{align}\label{lx to lu}
\nonumber l_x&=\dfrac{4\ell\sin{\dfrac{\l_\phi}{2}}}{\cos{\dfrac{l_\phi}{2}}+\cos{\dfrac{l_u}{2\ell}}},\\
l_t&=\dfrac{4\ell\sin{\dfrac{\l_u}{2\ell}}}{\cos{\dfrac{l_\phi}{2}}+\cos{\dfrac{l_u}{2\ell}}},
\end{align} 
then the equations for the spacelike curve \eqref{S-L in poincare} transforms into 
\begin{align}\label{spacelike BMS RT}
\nonumber & r=\dfrac{\ell\cos\left(\dfrac{l_u}{2\ell}\right)\sin\left(\frac{u}{\ell}\right)}{\cos\left(\dfrac{l_\phi}{2}\right)\cos(\phi)-\cos\left(\dfrac{l_u}{2\ell}\right)\cos(\dfrac{u}{\ell})},\\
& r^2\sin^2(\phi)\sin^2\left(\dfrac{l_\phi}{2}\right)=\sin^2\left(\dfrac{l_u}{2\ell}\right)\left(\ell\cos\left(u\over\ell\right)-r\sin\left(\dfrac{u}{\ell}\right)\right)^2.
\end{align}
 and the equations for the timelike curve \eqref{T-L in poincare} is written as
\begin{align}\label{timelike BMS RT}
\nonumber & r=\dfrac{\ell\cos\left(\dfrac{l_\phi}{2}\right)\sin\left(\frac{u}{\ell}\right)}{\cos\left(\dfrac{l_u}{2\ell}\right)\cos(\phi)-\cos\left(\dfrac{l_\phi}{2}\right)\cos(\dfrac{u}{\ell})},\\
& r^2\sin^2(\phi)\sin^2\left(\dfrac{l_\phi}{2}\right)=\sin^2\left(\dfrac{l_u}{2\ell}\right)\left(\ell\cos\left(u\over\ell\right)-r\sin\left(\dfrac{u}{\ell}\right)\right)^2.
\end{align}
It is clear that the second equation is the same for both curves. The spacelike curve intersects the boundary at points  $\left(u=\pm\frac{\ell l_\phi}{2},\phi=\pm\frac{l_u}{2\ell}\right)$. It should be noted that the condition $l_x>l_t$ is converted to $\sin\dfrac{l_\phi}{2}>\sin{\dfrac{l_u}{2\ell}}$ using \eqref{lx to lu}, which causes the  interval  $-\dfrac{\ell l_\phi}{2}<u<\dfrac{\ell l_\phi}{2}$,  $-\dfrac{l_u}{2\ell}<\phi<\dfrac{l_u}{2\ell}$ to be timelike.

The flat space limit  for both the spacelike and timelike curves is well-defined, and for the spacelike curve \eqref{spacelike BMS RT}, this limit yields  the  equations,
\begin{align}\label{flat extremal curve}
\nonumber &r=\dfrac{u}{\cos\phi\cos\left({l_\phi\over2}\right)-1},\\
&r^2\sin^2(\phi)\sin^2\left(\frac{l_\phi}{2}\right)=\dfrac{l_u^2}{4}.
\end{align}
 which are precisely the spacelike curve given in \cite{Jiang:2017ecm} in the framework of flat/CCFT as the holographic dual of entanglement entropy 
for the subsystem $-\dfrac{ l_\phi}{2}<\phi<\dfrac{ l_\phi}{2}$,  $-\dfrac{l_u}{2}<u<\dfrac{l_u}{2}$. The flat space limit of timelike curve \eqref{timelike BMS RT} results in
\begin{align}\label{flat Timelike }
\nonumber &r=\dfrac{u\cos\left({l_\phi\over2}\right)}{\cos\phi\ -\cos\left({l_\phi\over2}\right)},\\
&r^2\sin^2(\phi)\sin^2\left(\frac{l_\phi}{2}\right)=\dfrac{l_u^2}{4}.
\end{align}
This new timelike curve in the flat spacetime has not been studied in articles so far, and we will subsequently introduce it as the holographic dual for the imaginary part of the $\text{CCFT}$ pseudo-entropy. 

The equations \eqref{spacelike BMS RT} and \eqref{timelike BMS RT} for spacelike and timelike curves can be rewritten as follows. For the spacelike curve \eqref{spacelike BMS RT}, we have:

\begin{align}\label{eo new cureve rewritten}
\nonumber &\cos \phi=\dfrac{\cos\left(\dfrac{l_u}{2\ell}\right)\left(r\cos\dfrac{u}{\ell}+\ell\sin\dfrac{u}{\ell}\right)}{r\cos\dfrac{l_\phi}{2}},\\
& \cos^2\phi=\cos^2\dfrac{l_u}{2\ell}+\dfrac{\ell^2\sin^2\dfrac{l_u}{2\ell}\cos^2\dfrac{l_u}{2\ell}}{r^2\left(\sin^2\dfrac{l_u}{2\ell}-\sin^2\dfrac{l_\phi}{2}\right)},
\end{align}
and the timelike curve \eqref{timelike BMS RT} can also be rewritten as 
\begin{align}\label{r min and max}
\nonumber &\cos \phi=\dfrac{\cos\left(\dfrac{l_\phi}{2}\right)\left(r\cos\dfrac{u}{\ell}+\ell\sin\dfrac{u}{\ell}\right)}{r\cos\dfrac{l_u}{2\ell}},\\
& \cos^2\phi=\left(\dfrac{\cos^2\dfrac{l_\phi}{2}\sin^2\dfrac{l_\phi}{2}-\cos^2\dfrac{l_\phi}{2}\sin^2\dfrac{l_u}{2\ell}}{\cos^2\dfrac{l_\phi}{2}\sin^2\dfrac{l_\phi}{2}-\cos^2\dfrac{l_u}{2\ell}\sin^2\dfrac{l_u}{2\ell}}    \right)\left(1- \dfrac{\ell^2\sin^2\dfrac{l_u}{2\ell}}{r^2\left(\sin^2\dfrac{l_u}{2\ell}-\sin^2\dfrac{l_\phi}{2}\right)} \right)
\end{align}

\begin{figure}[h]
\centering\includegraphics[scale=1,width=110mm]{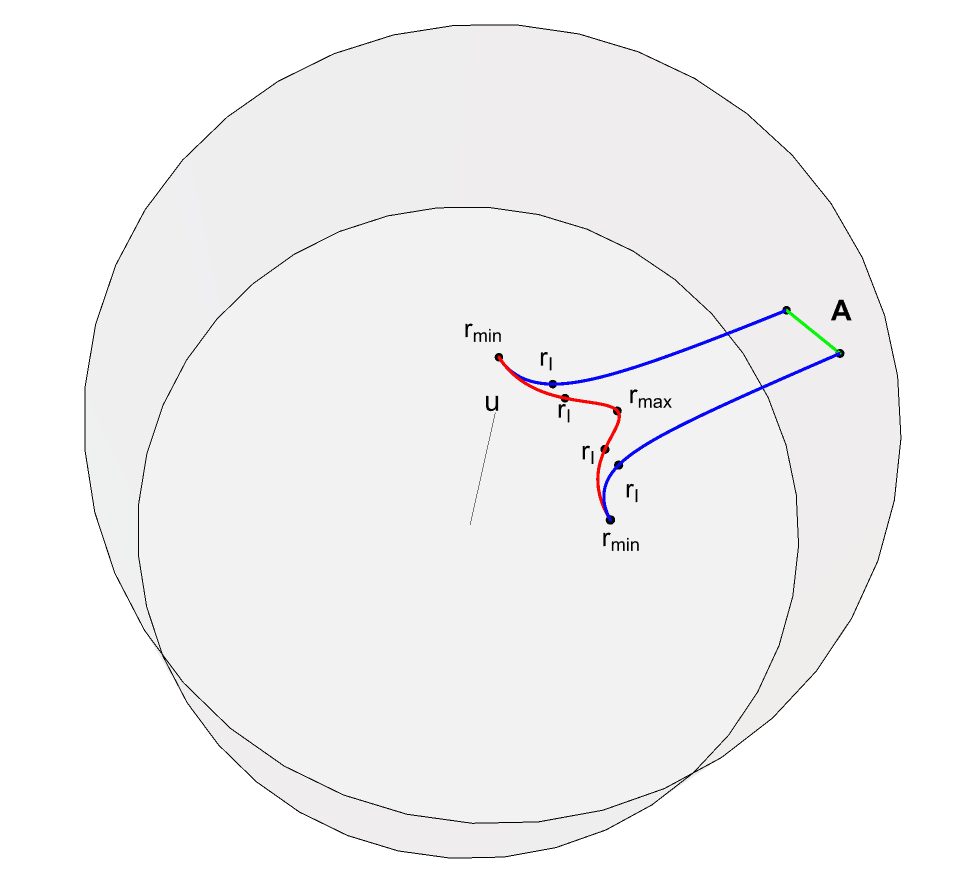}
\caption{ Spacelike and timelike extremal curves in BMS-AdS coordinate which are holographic dual of timelike entanglement entropy. Blue curve is spacelike and red curve is timelike. $A$ (green line) is timelike interval at the boundary.}\label{BMS adS fig}
\end{figure}

In BMS coordinates, the timelike curve is also connected to the two branches of the spacelike curve (Figure \ref{BMS adS fig}). The connection point has the minimum radius, denoted by $ r_{\text{min}} $. Using the second equation for both spacelike \eqref{eo new cureve rewritten} and timelike curves \eqref{r min and max}, $ r_{\text{min}} $ is obtained as
\begin{equation}\label{def of rmin}
r_{min}=\dfrac{\ell\sin\dfrac{l_u}{2\ell}}{\sqrt{\sin^2\dfrac{l_\phi}{2}-\sin^2\dfrac{l_u}{2\ell}}}.
\end{equation}

Furthermore, from the second equation of the timelike curve \eqref{r min and max}, it is derived that this curve can extend inward from the AdS boundary up to a certain radius, denoted by $r_{\text{max}}$, which is obtained as 
\begin{equation}\label{rmax}
r_{max}=\dfrac{\ell\cos\dfrac{l_\phi}{2}}{\sqrt{\sin^2\dfrac{l_\phi}{2}-\sin^2\dfrac{l_u}{2\ell}}}.
\end{equation}
Using the definition of $ r_{\text{min}} $ \eqref{def of rmin}, we can simplify the equation of the spacelike curve \eqref{eo new cureve rewritten} as 
\begin{align}\label{most simplified }
\nonumber &\cos \phi=\dfrac{\cos\left(\dfrac{l_u}{2\ell}\right)\left(r\cos\dfrac{u}{\ell}+\ell\sin\dfrac{u}{\ell}\right)}{r\cos\dfrac{l_\phi}{2}},\\
& \cos(\phi)=\dfrac{\cos\dfrac{l_u}{2\ell}}{r}\sqrt{r^2-r_{min}^2}.
\end{align}
By combining these two equations, we can eliminate $ \phi $ and obtain, 
\begin{equation}\label{u in terms of r}
r\cos\frac{u}{\ell}+\ell\sin\dfrac{u}{\ell}=\cos\dfrac{l_\phi}{2}\sqrt{r^2-r_{min}^2}.
\end{equation}
From this equation, it is evident that $u$ has an extremum value at radius $r = r_I $, where 
\begin{equation}\label{def of rstar}
r_I=\dfrac{\ell\sin\dfrac{l_u}{2\ell}\cos\dfrac{l_u}{2\ell}}{\sin^2\dfrac{l_\phi}{2}-\sin^2\dfrac{l_u}{2\ell}}.
\end{equation}
\subsection{ Length of timelike and spacelike curves and taking their flat-space limit}
In this subsection, we intend to calculate the length of the timelike and spacelike curves obtained in the previous subsection in BMS coordinates and then take their flat-space limit. Calculating the length of the curves is also more convenient in Poincaré coordinates. Therefore, instead of working with equations \eqref{eo new cureve rewritten} and \eqref{r min and max}, we proceed with the equations of the extremal curves in Poincaré coordinates, i.e., equations \eqref{S-L in poincare} and \eqref{T-L in poincare}.

Using the equations of the spacelike curve \eqref{S-L in poincare}, the line element \eqref{BMS-global} for this curve is obtained as 
\begin{align}\label{linesl}
   ds_{sl}=\dfrac{\ell A dz}{z\sqrt{z^2+A^2}},    
  \end{align}
and for the timelike curve \eqref{T-L in poincare}, the line element  is 
\begin{align}\label{linetl}
     ds_{tl}=\dfrac{i\ell A dz}{z\sqrt{z^2-A^2}},   
  \end{align}
where
\begin{equation}\label{Def of A}
  A=\dfrac{\sqrt{l_x^2-l_t^2}}{2}.
  \end{equation}
Now we can again use the coordinate transformation \eqref{coordinate change} and express the line elements \eqref{linesl} and \eqref{linetl} in BMS coordinates. We have
\begin{equation}\label{line in bms ads}
  ds_{sl}=\dfrac{-\ell d\Delta}{\sqrt{\Delta^2+\dfrac{4\ell^4}{A^2}}},\qquad ds_{tl}=\dfrac{-i\ell d\Delta}{\sqrt{\dfrac{4\ell^4}{A^2}-\Delta^2}} .
  \end{equation} 
Using equations \eqref{coordinate change}, \eqref{u in terms of r} , and \eqref{def of rmin}, we see that at $ r = r_{\text{min}} $, $ \Delta = 0 $. Therefore, integrating from an arbitrary initial radius such as $r = r_i $ (equivalent to $ \Delta = \Delta_i $) to $r = r_{\text{min}} $ (or $ \Delta = 0 $) will yield the following lengths for the spacelike and timelike curves:
\begin{equation}\label{Length of SL}
      L_i^{sl}=2\ell\sinh^{-1}\left(\dfrac{A\Delta_i}{2\ell^2}\right),
      \end{equation} 
\begin{equation}\label{Length of TL}
      L_i^{tl}=2i\ell\sin^{-1}\left(\dfrac{A\Delta_i}{2\ell^2}\right).
      \end{equation}       
If we take the radius $ r_i $ to be infinite (i.e., covering the entire length of the spacelike curve), then this length divided by $ 4G $ (where $ G $ is Newton's constant) will yield the real part of the pseudo-entropy in \cite{Doi:2022iyj, Doi:2023zaf}. 

For the timelike part, $ r_i $ cannot be larger than $ r_{\text{max}} $. If we calculate the length of this curve from $ r_{\text{max}} $ to $ r_{\text{min}}$ (i.e., the entire length of the timelike curve), the value $ i\pi\ell$ is obtained, which corresponds to the imaginary part of the pseudo-entropy in \cite{Doi:2022iyj, Doi:2023zaf}. 

\begin{figure}[h]
\centering\includegraphics[scale=1,width=110mm]{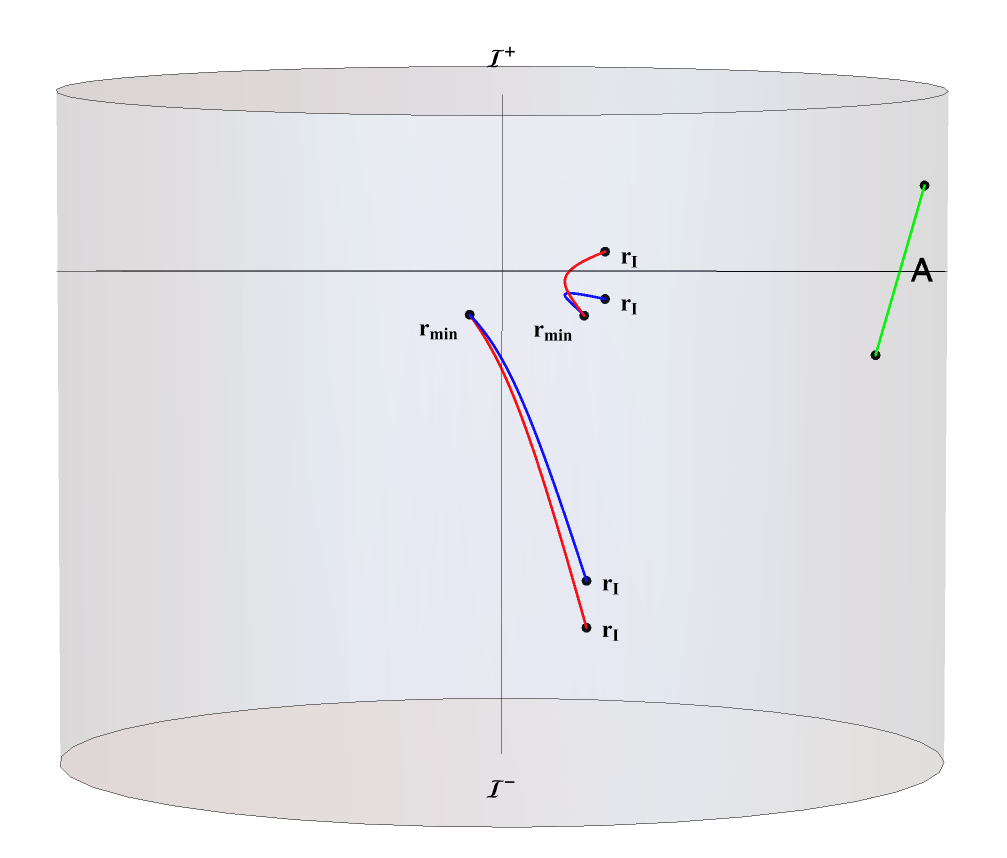}
\caption{ Extremal curves in flat spacetimes written in BMS coordinate. Spacelike curve is blue and timelike  curve is red. The length of these curves are proportional to the pseudo-entropy of an interval in CCFT.  }\label{flat fig}
\end{figure}

Both of these values lack a well-defined flat-space limit. Therefore, following the approach in  \cite{Fareghbal:2024lqa}, instead of considering the entire length, we only compute the portion from the junction point of the timelike and spacelike curves (i.e., $ r = r_{\text{min}} $) to $ r = r_I $, where the retarded time becomes extremal. Here, $r = r_I$ corresponds to $ \Delta = \Delta_I $, and we have:
\begin{equation}\label{def of delta star}
      \Delta_I=\dfrac{\ell\sin\dfrac{l_u}{2\ell}\cos\dfrac{l_\phi}{2}}{\cos\dfrac{l_u}{2\ell}-\cos\dfrac{l_\phi}{2}}
      \end{equation}  
Therefore, the lengths of the spacelike and timelike portions collectively lead to the following general expression:
\begin{equation}\label{final lentgh before flat-space limit for ads}
      L=2\ell\sinh^{-1}\left(\dfrac{\sin\dfrac{l_u}{2\ell}\cos\dfrac{l_\phi}{2}}{\sqrt{\cos^2\dfrac{l_u}{2\ell}-\cos^2\dfrac{l_\phi}{2}}}\right)+2i\ell\sin^{-1}\left(\dfrac{\sin\dfrac{l_u}{2\ell}\cos\dfrac{l_\phi}{2}}{\sqrt{\cos^2\dfrac{l_u}{2\ell}-\cos^2\dfrac{l_\phi}{2}}}\right)
      \end{equation}
The flat-space limit ($\ell \to \infty $) of this expression is well-defined, and we have:
\begin{equation}\label{Lflat}
L_{flat}=\left(1+i\right)\dfrac{l_u\cos\frac{l_\phi}{2}}{\sin \frac{l_\phi}{2}},
\end{equation}
The real part of this relation divided by $4G$ corresponds to the entanglement entropy expression in CCFT \cite{Bagchi:2014iea}. However, its imaginary part is novel. The spacelike and timelike extremal curves in flat spacetime are depicted in (figure \ref{flat fig}). The spacelike curves are the same as \cite{Jiang:2017ecm} but the timelike ones are new.
 If we follow the same approach as in references \cite{Doi:2022iyj, Doi:2023zaf}, we must accept that the entanglement entropy for CCFT is essentially a pseudo-entropy, comprising both real and imaginary parts. Our holographic calculation suggests that the imaginary part should be derived from the imaginary part of $L/4G$. A direct computation of this expression in CCFT remains an open problem that can be addressed in future studies.

\section{Pseudo-entropy in CCFT via dS holography}
The dictionary for flat/CCFT correspondence  can be derived not only from AdS/CFT but also from ds holography. In this context,  the holographic dual for pseudo-entropy is defined as the length of spacelike curves in asymptotically dS spacetimes for the real part, and the length of timelike curves for the imaginary part \cite{Doi:2022iyj, Doi:2023zaf}. We expect that by taking an appropriate limit of these curves, we can obtain the equivalent quantity in CCFT within the flat/CCFT framework.

\begin{figure}[h]
\centering\includegraphics[scale=1,width=110mm]{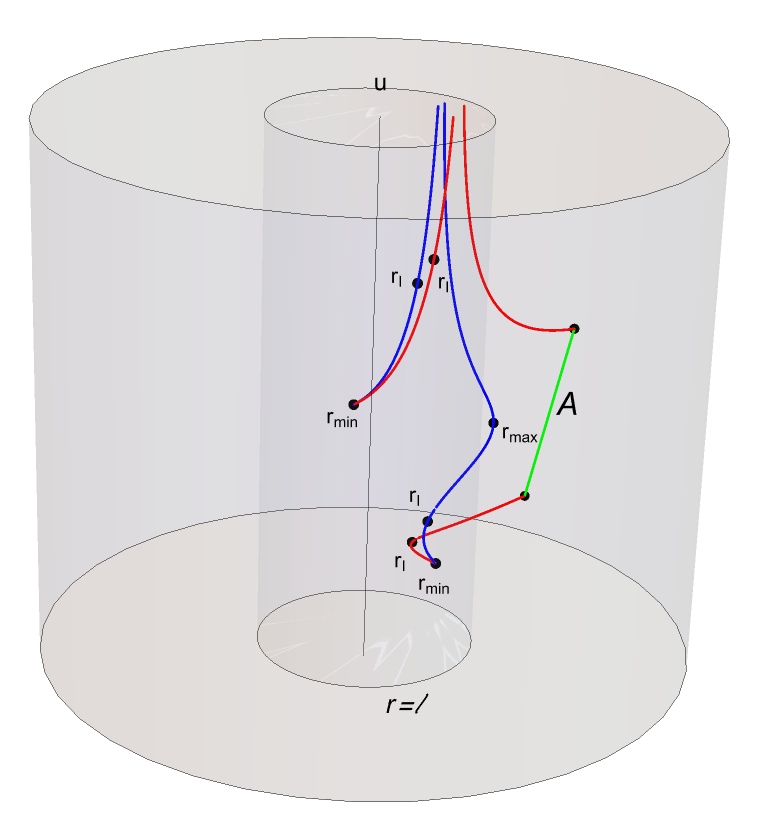}
\caption{ Spacelike (blue) and timelike (red) extremal curves in dS$_3$ which are holographic dual of pseudo-entropy. $A$ (green)  is an arbitrary interval in  CFT.    }\label{BMS dS fig}
\end{figure}

To this end, we begin with dS$_3$ space in BMS coordinates, which has the following metric:
\begin{equation}\label{ds in BMS}
ds^2=-\left(1-\dfrac{r^2}{\ell^2}\right)du^2-2du dr+r^2 d\phi^2,
\end{equation}
where $\ell$ is the radius of dS spacetime. This spacetime has a cosmological horizon located at $r=\ell$. By comparing the metrics of AdS$_3$ and dS$_3$ in BMS coordinates, we find that they can be transformed into each other by replacing $\ell$ with $i\ell$. Therefore, instead of directly deriving curves with a well-defined flat-space limit, we begin with the curves \eqref{spacelike BMS RT} and \eqref{timelike BMS RT} and apply the transformation $\ell\to i\ell$. The result is that the spacelike curve in dS$_3$ becomes,
\begin{align}\label{new extremal curve ds }
\nonumber & r=\dfrac{\ell\cosh\left(\dfrac{l_u}{2\ell}\right)\sinh\left(\frac{u}{\ell}\right)}{\cos\left(\dfrac{l_\phi}{2}\right)\cos(\phi)-\cosh\left(\dfrac{l_u}{2\ell}\right)\cosh(\dfrac{u}{\ell})},\\
& r^2\sin^2(\phi)\sin^2\left(\dfrac{l_\phi}{2}\right)=\sinh^2\left(\dfrac{l_u}{2\ell}\right)\left(\ell\cosh\left(u\over\ell\right)+r\sinh\left(\dfrac{u}{\ell}\right)\right)^2,
\end{align} 
and the timelike curves will be obtained as
\begin{align}\label{new extremal curve ds timekile }
\nonumber & r=\dfrac{\ell\cos\left(\dfrac{l_\phi}{2}\right)\sinh\left(\frac{u}{\ell}\right)}{\cosh\left(\dfrac{l_u}{2\ell}\right)\cosh(\phi)-\cos\left(\dfrac{l_\phi}{2}\right)\cos(\dfrac{u}{\ell})}, \\
& r^2\sin^2(\phi)\sin^2\left(\dfrac{l_\phi}{2}\right)=\sinh^2\left(\dfrac{l_u}{2\ell}\right)\left(\ell\cosh\left(u\over\ell\right)+r\sinh\left(\dfrac{u}{\ell}\right)\right)^2.
\end{align} 
The diagram corresponding to these curves is plotted in figure \ref{BMS dS fig}. As can be seen, unlike AdS$_3$, in this case the timelike curve extends to timelike infinity and connects to two endpoints of an interval in the dual CFT. This curve consists of two branches, one of them crosses the cosmological horizon and extends inward to the radius $r = r_{\text{min}}$ where
\begin{equation}\label{def of rmin ds}
r_{min}=\dfrac{\ell\sinh\frac{l_u}{2\ell}}{\sqrt{\sinh^2\dfrac{l_u}{2\ell}+\sin^2\dfrac{l_\phi}{2}}},
\end{equation}
 and connects to the spacelike curve. As shown in  figure 3, the spacelike curve also has two branches. One end of these branches connects to the timelike curve at $r_{\text{min}} $, and after crossing the cosmological horizon, at radius $r_{\text{max}}$ where
\begin{equation}\label{def of rmax}
r_{max}=\dfrac{\ell\cosh\dfrac{l_u}{2\ell}}{\sqrt{\sinh^2\dfrac{l_u}{2\ell}+\sin^2\dfrac{l_\phi}{2}}}.
\end{equation}
 the two branches connect to each other. The flat-space limit of curves \eqref{new extremal curve ds } and \eqref{new extremal curve ds timekile } is well-defined, and the spacelike part leads to equations for curves that were introduced in \cite{Jiang:2017ecm} as the dual to entanglement entropy in CCFT. However, the timelike curve has not been investigated within the flat/CCFT framework. Furthermore, we observe that the flat-space limit of curves \eqref{spacelike BMS RT} and \eqref{timelike BMS RT} in AdS space, and curves \eqref{new extremal curve ds } and \eqref{new extremal curve ds timekile } in dS space, yields the same result in flat space.

 From the equation of the spacelike curve \eqref{new extremal curve ds }, we can find the radius at which the retarded time $u$ becomes extremal. If we denote this radius as $r_I$, condition $\dfrac{du}{dr}\Big|_{r=r_I}=0$  leads to the following value for $r_I$:
 \begin{equation}\label{def of r star ds}
r_I=\dfrac{r_{min}r_{max}}{\ell}.
\end{equation}
 Additionally, it is not difficult to see that $r_I < \ell $, meaning this radius and the entire portion $r_{min}\leq r\leq r_I$ of the spacelike curve lie inside the cosmological horizon.

To obtain the length of the curves, one can also apply the transformation $\ell \to i\ell $ to the calculations in the previous section. The length of the portion of the curves within interval $r_{min}\leq r\leq r_I$, obtained through this transformation from relation \eqref{final lentgh before flat-space limit for ads}, is as follows:
\begin{equation}\label{L star}
 L_I=2\ell\sin^{-1}\dfrac{\sinh\frac{l_u}{2\ell}\cos\frac{l_\phi}{2}}{\sqrt{\sinh^2\dfrac{l_u}{2\ell}+\sin^2\frac{l_\phi}{2}}} + 2i\ell\sinh^{-1}\dfrac{\sinh\frac{l_u}{2\ell}\cos\frac{l_\phi}{2}}{\sqrt{\sinh^2\dfrac{l_u}{2\ell}+\sin^2\frac{l_\phi}{2}}}
 \end{equation} 
The flat-space limit of this equation, similar to relation \eqref{final lentgh before flat-space limit for ads}, leads to the following result:
\begin{equation}
\dfrac{1}{4 G}\lim_{\ell\to\infty} L_I=\left(1+i\right)\dfrac{l_u}{4G}\cot\frac{l_\phi}{2},
\end{equation}
The real part corresponds to the entanglement entropy in CCFT \eqref{ee cylinder} with $ c_L = 0 $ and $ c_M = \frac{3}{G} $, while the imaginary part indicates that this entropy is essentially a pseudo-entropy.

 \section{Summary and future directions }
In this paper, using flat/CCFT correspondence, we introduced pseudo-entropy for CCFTs that are dual to three-dimensional Minkowski spacetime. The pseudo-entropy is inherently a complex quantity whose real part is dual to the length of spacelike extremal curves in the three-dimensional flat spacetime, while its imaginary part is obtained from the length of timelike extremal curves in the same spacetime. Our approach in this work began with (A)dS/CFT  correspondences, and by taking the flat-space limit of the extremal curves dual to the pseudo-entropy in the CFT. Starting from both (A)dS/CFT  correspondences, we arrived at the same curves in Minkowski spacetime. The spacelike curves had previously been introduced in \cite{Jiang:2017ecm}, but the timelike curves constitute one of the main results of this paper.

Our main calculation in this paper was in the bulk, where we showed that both spacelike and timelike extremal curves in AdS and dS spacetimes have a well-defined flat space limit. However, the flat space limit from the intervals that these curves are related to their pseudo-entropy is not well-defined, and indeed, our method did not fully clarify  the  interval that the pseudo-entropy corresponds to it.

Since the spacelike curve was related to the entropy of the interval introduced in \cite{Jiang:2017ecm}, we also associated the timelike curve obtained in this paper which was also connected to the spacelike curve, with the imaginary part of the pseudo-entropy of the same interval. In fact, according to the flat/CCFT correspondence, this interval should be at the null infinity of flat spacetime. In contrast, similar intervals, whose spacelike and timelike curves in AdS and dS had flat space limits leading to the curves of this interval, should be defined on the spacelike boundary of AdS and the timelike infinity of dS. Since, in the flat space limit, none of these hyper-surfaces in AdS, dS, and flat spacetimes transform into one another, we should not expect that the interval introduced in \cite{Jiang:2017ecm} can be obtained from the flat limit of similar intervals in asymptotically AdS and dS spacetimes. Therefore, because the flat-space limit is not well-defined on the boundary, we attribute the curves we obtained in flat spacetime to the pseudo-entropy of the interval introduced in reference \cite{Jiang:2017ecm}.

Another point that compels us to accept the pseudo-entropic nature of the quantity introduced for entanglement entropy in CCFT is that, alongside efforts deriving the flat/CCFT dictionary from AdS/CFT (see also new papers \cite{Banerjee:2024ldl}-\cite{Kulkarni:2025qcx}), this correspondence should also be obtainable from dS holography. Two main approaches for the holography of asymptotically dS spacetimes are given in \cite{Strominger:2001pn} and \cite{Alishahiha:2004md} . In \cite{Strominger:2001pn},  the CFT dual to asymptotically dS spacetimes  is non-unitary, and we deal with pseudo-entropy rather than entanglement entropy. Therefore, it is expected that after taking the limit, pseudo-entropy would be obtained. The pseudo-entropy we encounter in CCFT is not of the same nature as the timelike entanglement entropy we had in Ads/CFT, since the inherently ultra-relativistic nature of CCFT does not allow intervals to be distinguished in these two ways. We note that our calculation in this paper  was in the bulk dS spacetimes which is the same for both of the proposals of dS holography in \cite{Strominger:2001pn} and \cite{Alishahiha:2004md}. Interpretation of results as the holographic dual of CFT pseudo-entropy is based on  \cite{Strominger:2001pn} and \cite{Doi:2022iyj, Doi:2023zaf}. However, study of this problem in the context of dS/dS correspondence  \cite{Strominger:2001pn} is an interesting problem which may be utilizable by  following recent progresses in  \cite{Geng:2019bnn}-\cite{Geng:2021wcq}.

The possibility of defining pseudo-entropy for CCFTs could serve as evidence for the non-unitarity of these theories \cite{Hao:2025ocu}. This claim has not yet been proven for CCFTs without resorting to holographic methods and remains an open problem that we intend to address in future works. Furthermore, completing the calculations presented in this paper for other asymptotically flat spacetimes in three dimensions, as well as generalizing the framework to higher dimensions(particularly four) will be among our future research directions.

\subsubsection*{Acknowledgments}
This work is based upon research funded by Iran National Science Foundation (INSF) under project No 4047971.


\end{document}